\documentclass[a4paper,11pt]{article}
\usepackage{jinstpub} 
\usepackage{lineno}
\usepackage{bm}
\usepackage{siunitx}
\usepackage{subfigure}
\usepackage[]{caption}
\usepackage{multirow}
\usepackage{comment}

\title{\boldmath First complete characterization of an X-Ray tube through combined measurements and Geant4 simulations}

\author[b]{A.~Braghieri,}
\author[a,b]{M.~Brunoldi,}
\author[a,b]{S.~Calzaferri,}
\author[a,b]{M.~Pelliccioni,}
\author[b]{P.~Salvini,}
\author[b,1]{A.~Tamigio, \note{Corresponding author.}}
\author[a,b]{I.~Vai,}
\author[a,b]{P.~Vitulo}

\affiliation[a]{Dipartimento di Fisica, Università di Pavia, Via Bassi 6, 27100, Pavia, Italy}
\affiliation[b]{Istituto Nazionale di Fisica Nucleare Sezione di Pavia, Via Bassi 6, 27100, Pavia, Italy}

\emailAdd{alessandro.tamigio01@universitadipavia.it}

\abstract{X-ray tubes are sources of X-rays used in various fields, ranging from radiographic imaging in medical physics to the characterization of detectors in particle physics. This article presents a method for the complete characterization of the \textit{Mini-X2 X-Ray tube} from AMETEK, an X-ray source later used to characterize gas detectors in laboratory. Using data provided by the manual, we derive key operational parameters of the tube, such as the photon emission rate and the relationship between emissive power and supplied current. Based on this characterization, we determine the air dose rate as well as the absorbed dose rate within a given volume at various distances from the source. We derive this dose experimentally, theoretically and through simulations and find good agreement between these. To enable safe operation in a laboratory environment and to support experimental comparisons, a shielding system was developed. Finally, it will be illustrated how the shielding containing the Mini-X2 has been designed, exploiting the simulation power of the Geant4 software.}


\begin{document}
\maketitle
\flushbottom

\section{Operation of an X-ray tube}
\label{sec:intro}

\begin{figure}[htbp]
\centering
\includegraphics[width=.7\textwidth]{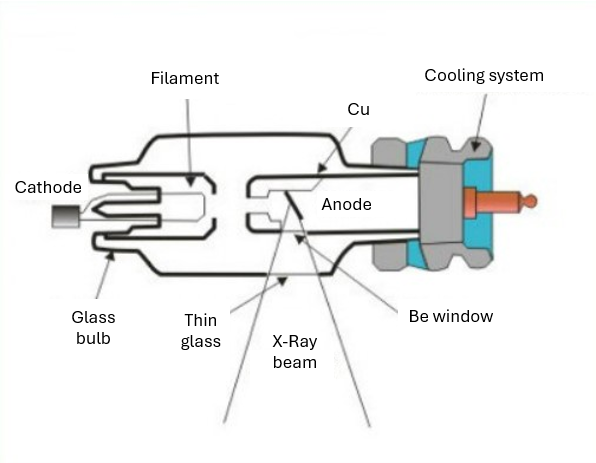}
\caption{Schematic X-ray tube representation~\cite{ref3}.\label{fig:tubo a raggi-X}}
\end{figure}

To understand the operation of an X-ray tube, it is convenient to start with its geometric description. As shown in Figure  \ref{fig:tubo a raggi-X}, an X-ray tube consists of two electrodes: the anode and the cathode placed in vacuum in a glass bulb.

The cathode usually consists of one or two tungsten filaments. The anode consists of a structure of different materials depending on the type of X-ray tube (Ag, Au, Rh, or W): when struck by the accelerated electrons, the anode emits X-rays through  \textit{Bremsstrahlung} or through characteristic radiation.

Electrons are emitted from the filament itself by thermoionic effect when the cathode filament is heated by currents of a few $\mu A$. At this point, the electronic cloud formed close to the cathode is accelerated toward the anode by the application of a positive voltage (typically some $kV$). This process generates the \textit{tube current}. Since a vacuum is created in the space between anode and cathode, electrons do not collide with gas molecules in passing through the tube itself. At the surface of the anode, the maximum kinetic energy acquired by these electrons is $E_{k} = \Delta V e$, where $\Delta V$ represents the difference in potential between anode and cathode and $e$ the electron charge. When electrons collide with the anode about 99\% of the energy is converted into heat and the 1\% into X-radiation. For this reason, as shown in figure \ref{fig:tubo a raggi-X}, the structure of an X-ray tube provides a cooling system that allows for a dissipation of the generated heat. Since X-radiation is produced isotropically, a shielding system that reduces the angle of the radiation emission is provided. From Figure \ref{fig:tubo a raggi-X}, it is possible to notice the presence of a beryllium (Be) window and thin glass are represented, which allow X-radiation to exit only for a given angle \cite{ref1}. 

\begin{figure}[htbp]
\centering
\includegraphics[width=.6\textwidth]{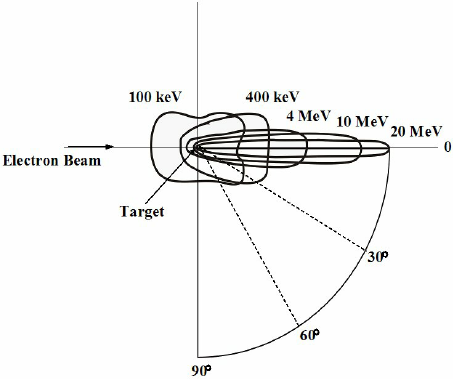}
\caption{Longitudinal angular distribution of X-rays as a function of the energy of electrons impinging on the anode~\cite{ref4}.}
\label{fig:Distribuzione angolare vs energia}
\end{figure}

The direction of propagation of the produced X beam depends  on the energy of the beam itself as shown in Figure \ref{fig:Distribuzione angolare vs energia}. The target of the anode is a thick layer that indeed can reabsorb the emitted electron. This leads to a more dispersive X-ray angular distribution if the photon energy is low. Otherwise, if the beam collides with the target at an energy of at least a few MeV, X-rays will be emitted mainly along the direction of the electron beam.

Finally, it is interesting to introduce two characteristics of the X-ray beam produced: the \textit{beam intensity} and the \textit{beam quality} \cite{ref5}. Beam intensity refers to the number of photons released by the X-ray tube when it is in operation. It depends on the number of electrons striking the anode, and thus on the current applied to the tube. It will be shown that there is direct proportionality between the current and the number of photons released. The applied voltage is closely related to beam quality and it refers to the energy of photons. A higher quality beam corresponds to photons with higher energy and thus to X-rays that are able to pass through thicker medium. The formula $E_{k} = e \Delta V$ shows the relationship between electron energy and potential difference between the electrodes. It is expected that a change in applied voltage will lead to a change in the X-ray energy spectrum.

\section{Energy spectrum of emitted X-rays}

The interaction of electrons with the target leads to the production of photons through two specific processes: \textit{Bremsstrahlung radiation} and \textit{characteristic radiation}. Both processes contribute to the formation of the energy spectrum of photons exiting the X-ray tube.

\section{Characterization of the Mini-X2 X-Ray tube}
\label{sec:intro}

\begin{figure}[htbp]
\centering
\includegraphics[width=.6\textwidth]{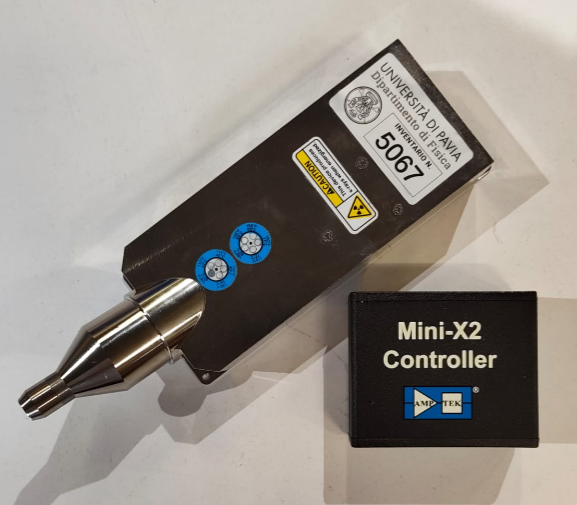}
\caption{Mini-X2 X-Ray tube from the University of Pavia \cite{ref7}.}
\label{fig:Mini-X2}
\end{figure}

We studied the features of the \textit{Mini-X2 X-ray tube}~\cite{ref7}. It consists of two components: the Mini-X2 X-ray tube module and the Mini-X2 controller (Figure \ref{fig:Mini-X2}). The first is the tube from which the produced X-rays come out, while the second is the controller that allows selection of both current and operating voltage.

To work safely with an X-ray source it is necessary to operate it only if the emitted photon flux is adequately shielded by absorbing material. Therefore, before using the Mini-X2, it is necessary to carry out a series of simulations with the \textit{Geant4} \cite{ref12} software that reproduces the experimental setup. In this way, appropriate shielding can be designed to protect the operator of the device.

For the simulation, a description as detailed as possible of the device is necessary, starting from the data provided in its manual. We obtained the following insight from the \textit{Geant4} study:

\begin{itemize}
    \item energy spectrum analysis;
    \item analysis of the photon flux emitted by the Mini-X2 at different applied currents;
    \item analysis of the angular distribution of emitted photons;
    \item calculation of absorbed dose in air at different distances;
    \item calculation of absorbed dose on an individual using shielding of different materials.
\end{itemize}

To reconstruct the procedures carried out for the analysis of Mini-X2, it is useful to list in Table \ref{tab:1} the information provided by the manual by which it was possible to characterize the source.

\begin{table}[htbp]  
    \centering 
    \caption{Mini-X2 X-Ray tube specifications~\cite{ref7}.} 
    \smallskip
    \begin{tabular}{l|c}
    \hline
       \textbf{Specifications}  &  \\
       \hline
       Anode material  &  Silver (Ag)\\
       Anode thickness  &  $\SI{0.8}{\mu m} \pm \SI{0.1}{\mu m}$\\
       Cathode material  &  Tungsten (W)\\
       Voltage applicable to the tube  &  from 0 to $\SI{50}{kV}$\\
       Current applicable to the tube  &  from $\SI{5}{\mu A}$ to $\SI{200}{\mu A}$\\
       Approximate dose rate  &  $\SI{1}{Sv/h}$ to $\SI{30}{cm}$ on the axis, $\SI{50}{kV}$, $\SI{80}{\mu A}$\\
       Output cone angle  &  $120$°\\
       \hline
    \end{tabular}   
    \label{tab:1}
\end{table}

\paragraph{Energy spectrum}
\label{sec:intro}

The energy spectrum of photons emitted by the Mini-X2 in the case of maximum voltage applied \cite{ref7} is composed by a continuous part due to Bremsstrahlung, with the typical linear tail at higher energies, and peaks due to the characteristic radiation of silver (Ag), caused by the transitions $K_\alpha$ e $K_\beta$. The maximum energy of the outgoing photons in the case of an applied voltage $\SI{50}{kV}$ results to be $\SI{50}{keV}$.

The energy values corresponding to the two characteristic peaks of Ag ($K_\alpha$ e $K_\beta$) can be obtained from the National Institute of Standards and Technology (NIST) database \cite{ref8} are given in Table \ref{tab:3}.

\begin{table}[htbp]   
    \centering  
    \caption{Silver (Ag) allowed transitions~\cite{ref8}}
    \smallskip
    \begin{tabular}{l|c}
    \hline
       \textbf{Transitions}  &  \textbf{Energy} \\
       \hline
       $K - L_{I}$  &  21.709 keV\\
       $K - M_{I}$  &  24.797 keV\\
       \hline
    \end{tabular}   
    \label{tab:3}
\end{table}

In Table \ref{tab:3} it can be seen that the most favorable transitions are characterized by radiation with energy around the values of \SI{22}{keV} and \SI{25}{keV}. A further consideration on the energy spectrum of the Mini-X2 concerns the quality of the beam. In fact, by changing the applied voltage value, different output spectra can be observed \cite{ref7}.

However, in the work presented the Mini-X2 was operated at \SI{50}{kV}, this implied that the outgoing spectrum always included the tail to 50 keV. Therefore, shielding design and simulations had to take into account the presence of more penetrating rays.

\paragraph{Numerical calculation of photons emitted per second}
\label{sec:intro}

By calculating the number of photons emitted per second from the Mini-X2 it is possible to quantify the particles leaving the source and incorporate this value within the simulations. In addition, considering the relationship between absorbed dose and flux, using the number of photons allowed the calculation of the dose at different distances from the Mini-X2.

The number of photons emitted per second was calculated from the dose rate provided by the manual and presented in Table \ref{tab:1}:

\begin{equation} 
\label{eq:3.4}
\dot D_{air} = 1.00 \pm 0.01 \hspace{0.1cm} Sv/h
\end{equation}
at \SI{30}{cm} from the longitudinal axis, operating at \SI{50}{kV}, with a current of $\SI{80}{\mu A}$ in which the uncertainty\footnote{The manual mentions an approximate dose rate.} on the dose was assumed to be 1\%.

The relationship linking the dose rate in air with the number of photons in that position is 

\begin{equation} 
\label{eq:2.13}
\dot D_{air} = 5.76 \cdot 10^{-4} \int_{E_{in}}^{E_{fin}} \!\!\! \frac{\phi_E}{\pi \cdot r^2} \hspace{0.04cm} E_{\gamma} \left( \frac{\mu_{en}}{\rho} \right)_{E}dE
\end{equation}
where $\mu_{en}/\rho$ represents the massive attenuation coefficient [$cm^2/g$], $r$ is the distance from the source, and $\phi_{E}$ is the number of photons of energy $E$ emitted by the Mini-X2 per second.

Placing these data within formula (\ref{eq:2.13}) yields the following:

\begin{equation} 
\label{eq:3.5}
10^3\hspace{0.1cm}mSv/h = 5.76 \cdot 10^{-4} \int_{E_{in}}^{E_{fin}} \!\!\! \frac{\phi_E}{\pi \cdot 30^2} \hspace{0.04cm} E_{\gamma} \left( \frac{\mu_{en}}{\rho} \right)_{E}dE
\end{equation}
in which

\begin{equation} 
\label{eq:3.6}
\phi_E = k_E \hspace{0.06cm}\phi
\end{equation}
where $\phi$ is the total number of photons produced by the source in units time, and $k_E$ is the weight of the energy channel $E$. It is therefore possible to rewrite (\ref{eq:3.5}) by taking $\phi$ out of the integral. Isolating this term gives:

\begin{equation} 
\label{eq:3.7}
\phi = 4.92 \cdot 10^9 \left[\int_{E_{in}}^{E_{fin}} \!\!\! k_E \hspace{0.04cm} E_{\gamma} \left( \frac{\mu_{en}}{\rho} \right)_{E}dE \right]^{-1}
\end{equation}
Having obtained the generic equation for the number of photons, it is necessary to solve two problems given by (\ref{eq:3.7}). The first concerns the functional form of the $E_{\gamma}$ spectrum, while the second is related to the value of the massive attenuation coefficient, whose dependence with energy is unknown. To get around the lack of knowledge of the functional form of the $E_{\gamma}$ spectrum and the analytical trend of $\mu_{en}$ as the energy varies, we resorted to an approximate process by discretizing the spectrum. This approach allows the rewriting of (\ref{eq:3.7}) in the form

\begin{equation} 
\label{eq:3.8}
\phi = 4.92 \cdot 10^9 \left[\sum_{E=E_{in}}^{E_{fin}} k_E \hspace{0.06cm} E_{\gamma} \hspace{0.06cm} \left(\frac{\mu_{en}}{\rho}\right)_E \right]^{-1}
\end{equation}

At this point, it is necessary to determine the value of the summation by working on the terms within it. From the spectrum of photons, it can be seen that the energy varies from the minimum value $E_{in} = 0$ to the maximum value $E_{fin} = 50\hspace{0.06cm}keV$, therefore both the extremes of the summation and the possible values of the term $E_i$ are known. 

Regarding the evaluation of the weight $k_i$ of the i-th energy channel it was necessary to proceed as follows. An image of the energy spectrum of the Mini-X2 was taken, divided into pixels\footnote{The image has been divided into 200 columns along the energy axis, so each channel corresponds to 4 columns of pixels.} and counted, channel by channel, the whole number of pixels present. Once the values were obtained, it was possible to calculate the weight of the individual channel as

\begin{equation} 
\label{eq:3.9}
k_i = \frac{channel \hspace{0.12cm} pixel \hspace{0.12cm} number}{total \hspace{0.12cm} pixel \hspace{0.12cm} number}
\end{equation}

\begin{figure}[htbp]
\centering
\includegraphics[width=.92\textwidth]{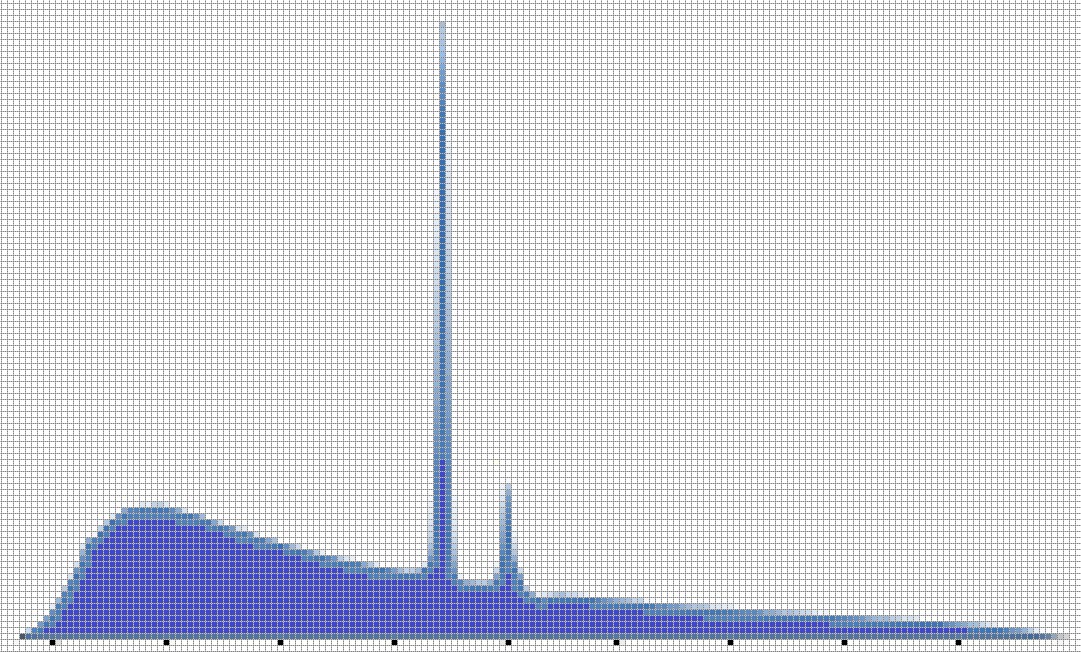}
\caption{Pixel division of the spectrum.}
\label{fig:tubo a raggi X}
\end{figure}

Because of this rough granularity, the square counting contains some inaccuracy. Specifically, for each energy channel, an error of $\pm 1$ was applied. Summing all channels yields

\begin{equation} 
\label{eq:3.10}
total \hspace{0.12cm} pixel \hspace{0.12cm} number \hspace{0.12cm} in \hspace{0.12cm} the \hspace{0.12cm} spectrum= 2347 \pm 46
\end{equation}

The next step in calculating the summation in Equation (\ref{eq:3.8}) is to evaluate the massive attenuation coefficient in air as a function of energy. Some of these values were available within the NIST table \ref{tab:10}, but those missing were calculated.

\begin{figure}[htbp]
\centering
\includegraphics[width=1\textwidth]{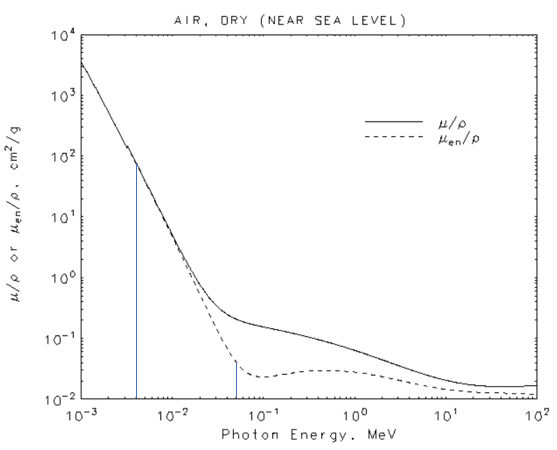}
\caption{Log-log scale plot of the attenuation coefficient of X-rays in air (solid line) and the mass attenuation coefficient (dashed line). The two vertical lines delimit the range of photon energies emitted by the Mini-X2~\cite{ref9}.}
\label{fig:Air NIST}
\end{figure}

\begin{table}[htbp]  
    \centering  
    \caption{$\mu_{en}/\rho$ values from NIST~\cite{ref9}.} 
    \smallskip
    \begin{tabular}{c|c|c}
    \hline
        \textbf{Energy (MeV)} & $\bm{\mu/\rho}$ $\bm{(cm^2/g)}$ &  $\bm{\mu_{en}}$/$\bm{\rho}$ $\bm{(cm^2/g)}$ \\ 
       \hline
       1.00000E-03  &  3.606E+03  &  3.599E+03  \\
       1.50000E-03  &  1.191E+03  &  1.188E+03  \\
       2.00000E-03  &  5.279E+02  &  5.262E+02  \\
       3.00000E-03  &  1.625E+02  &  1.614E+02  \\
       3.20290E-03  &  1.340E+02  &  1.330E+02  \\
       3.20290E-03  &  1.485E+02  &  1.460E+02  \\
       4.00000E-03  &  7.788E+01  &  7.636E+01  \\
       5.00000E-03  &  4.027E+01  &  3.931E+01  \\
       6.00000E-03  &  2.341E+01  &  2.270E+01  \\
       8.00000E-03  &  9.921E+00  &  9.446E+00  \\
       1.00000E-02  &  5.120E+00  &  4.742E+00  \\
       1.50000E-02  &  1.614E+00  &  1.334E+00  \\
       2.00000E-02  &  7.779E-01  &  5.389E-01  \\
       3.00000E-02  &  3.538E-01  &  1.537E-01  \\
       4.00000E-02  &  2.485E-01  &  6.833E-02  \\
       5.00000E-02  &  2.080E-01  &  4.098E-02  \\
       \hline
    \end{tabular}     
    \label{tab:10}
\end{table}

\begin{figure}[htbp]
\centering
\includegraphics[width=0.9\textwidth]{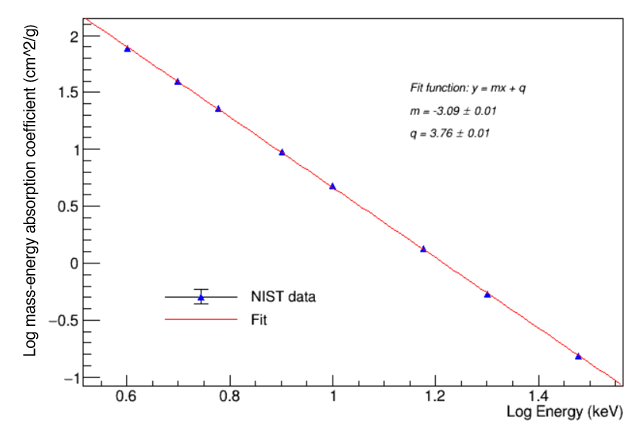}
\caption{Graph of $\mu_{en}/\rho$ values in log-log scale from NIST. The result of a linear fit to the values is superimposed.}
\label{fig:Fit dati NIST}
\end{figure}

Between $0$ and \SI{50}{keV}, it is observed that the $\mu_{en}/\rho$ trend is linear in a bilogarithmic scale, as expected. The region around \SI{50}{keV} does not follow this trend. Therefore, by fitting the data with a straight line, it is possible to derive the equation describing the bilogarithmic trend of the coefficient, and the fit and its parameters are shown in Figure \ref{fig:Fit dati NIST}.

\begin{figure}[htbp]
\centering
\includegraphics[width=0.9\textwidth]{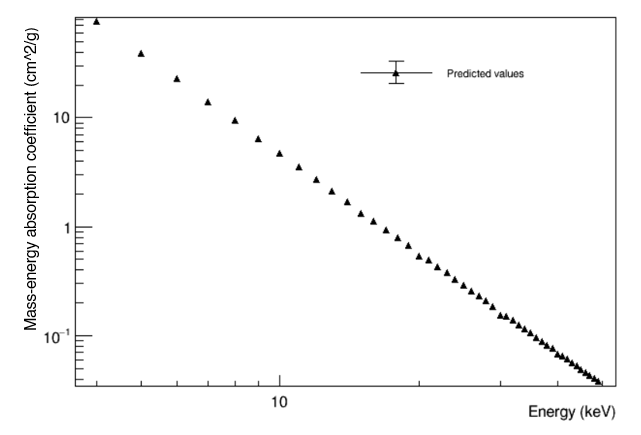}
\caption{Graph of $\mu_{en}/\rho$ values predicted from the fit to the NIST data.}
\label{fig:Coefficienti trovati}
\end{figure}

From this result we derived the equation 

\begin{equation} 
\label{eq:3.11}
\log\left (\frac{\mu_{en}}{\rho}\right) = -3.089\cdot\log(E)+3.756 
\end{equation}
It is then possible to predict the values of $\mu_{en}/rho$ for any value of energy between $0$ and \SI{50}{keV} using equation (\ref{eq:3.11}). Figure \ref{fig:Coefficienti trovati} contains the predictions obtained with this method. We find good compatibility between the predicted values and the ones provided by NIST for specific energy points showed in table \ref{tab:10}.

At this point, returning to formula (\ref{eq:3.8}), it is possible to substitute within the summation all the information about the 50 channels, obtaining the number of photons emitted per second by the Mini-X2 when applying a current of $\SI{80}{\mu A}$. Propagating the uncertainties as well, we obtain

\begin{equation} 
\label{eq:3.12}
\phi_{80\mu A} = (1.81 \pm 0.11) \cdot 10^{11} \hspace{0.12cm} \gamma/s
\end{equation}
From this number we derive, under the assumption of linearity, the amount of photons emitted for the maximum current of $\SI{200}{\mu A}$.

\begin{equation} 
\label{eq:3.13}
\phi_{200\mu A} = (4.53 \pm 0.29) \cdot 10^{11} \hspace{0.12cm} \gamma/s
\end{equation}
A clarification should be made about the results obtained. These data were derived theoretically from the approximate dose under the conditions in the manual. However, that dose is ideal and does not take into account the actual physical processes involving photons at lower energies. For example, the figure of \SI{1}{Sv/h} at \SI{30}{cm} distance encapsulates the contribution to the dose given even by those X-rays that physically lose all their energy after very few centimeters in air and therefore do not reach the \SI{30}{cm} distance at which the dose is calculated.

Therefore, the number of emitted photons found is correct, but what we expect to experimentally measure a lower dose, devoid of the contribution given by X-rays  that are attenuated by ionization on the path between the source and the detector. 

\paragraph{Calculation of the dose rate in air}
\label{sec:intro}

The theoretical calculation of the dose rate in air as a function of distance from the source allows for an estimate of it in the case where the Mini-X2 is used at maximum current ($\SI{200}{\mu A}$) and without taking into account the absorption of photons in in the air volume. The dose rate is given by the formula 

\begin{equation} 
\label{eq:3.14}
\dot D_{air} = 5.76 \cdot 10^{-4} \frac{\phi}{\pi \cdot r^2}\sum_{i=E_{in}}^{E_{fin}} k_i \hspace{0.06cm} E_i \hspace{0.06cm} \left(\frac{\mu_{en}}{\rho}\right)_i
\end{equation}
where $\phi$ is given by (\ref{eq:3.13})  and the distance is variable. The summation in the case where all energy channels between 0 and \SI{50}{keV} are considered, is $2.79 \cdot 10^{-2}$ $MeV \cdot cm^2/g$.

\paragraph{Comparison between predicted and experimental dose}
\label{sec:intro}

\begin{figure}[htbp]
\centering
\includegraphics[width=1\textwidth]{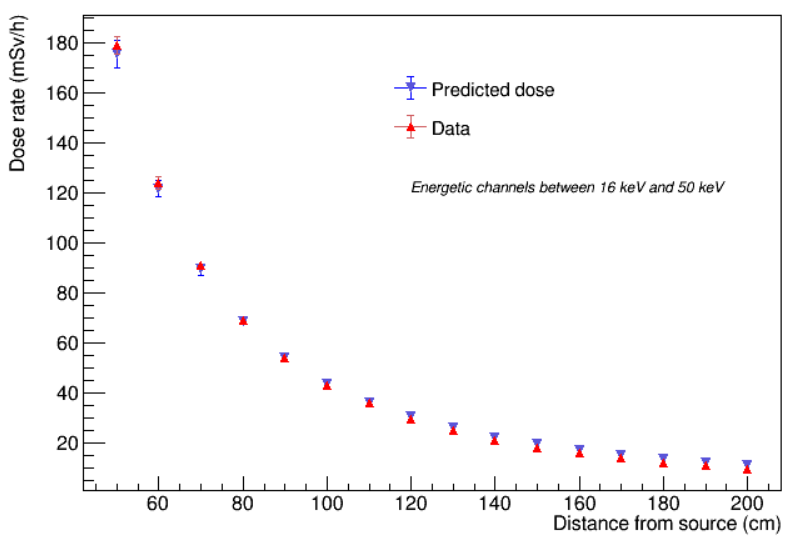}
\caption{Comparison between predicted and measured dose rate.}
\label{fig:Confronto tra dose teorica e dose sperimentale}
\end{figure}

To find out whether the number of emitted photons assumed in (\ref{eq:3.13}) is valid and whether the beam can really be considered isotropic, theoretical dose values at different distances, calculated through (\ref{eq:3.14}), and experimental data from measurements. The experimental measurements was carried out by varying the distance between a dosimeter and the Mini-X2 along the longitudinal axis of the X-ray tube. The dosimeter had a minimum threshold of sensitivity for photons at energy \SI{16}{keV}~\cite{ref10}. Therefore, for comparison, the theoretical dose was calculated by including in the (\ref{eq:3.14}) $E_{in}$ = \SI{16}{keV} and $E_{fin}$ = \SI{50}{keV}, so that only the contribution due to these channels was considered in the dose rate. In Figure \ref{fig:Confronto tra dose teorica e dose sperimentale} it is observed that the theoretical predictions are in perfect agreement with the experimental data thus demonstrating that $\phi_{200 \mu A}$ is a very good approximation for the number of photons emitted per second and that it is possible to consider the emission as isotropic.

\section{Simulations for the construction of the shield}
\label{sec:intro}

\paragraph{Geant4 simulation software}
\label{sec:intro}

Geant4 \cite{ref12} is one of the leading tools for simulating the passage of particles through matter. Using a Monte Carlo method, it is able to simulate the physical processes and interactions that particles undergo. It was used for two purposes:

\begin{itemize}
\item Assess through simulation the dose values studied in the previous section.
\item Simulate the shielding construction.
\end{itemize}
Only after simulating the construction of a shield around the Mini-X2 was Geant4 used to evaluate the dose absorbed by a person close to the source. 

\paragraph{Source simulation}
\label{sec:intro}

In this section, results obtained from simulations of the source by Geant4 will be shown; we expect the absorbed dose rate result to be in good agreement with the predicted and measured values.

\begin{figure}[htbp]
\centering
\includegraphics[width=.4\textwidth]{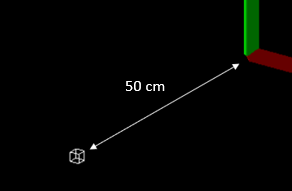}\label{fig:imm}
\qquad
\includegraphics[width=0.4\textwidth]{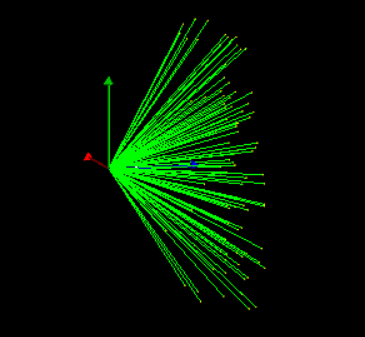}\label{fig:Fascio e detector Geant4}
\caption{Positioning of the detector at \SI{50}{cm} from the source (left). Simulation of $100$ photons emitted at 120° (right).\label{fig:i}}
\label{fig:Dosimentro a distanza 50 cm e fascio GPS}
\end{figure}

The simulation was carried out by entering these parameters: 
\begin{itemize}
\item $\phi$ = $4.5$ $\cdot$ $10^{8}$ $\gamma$ distributed according to the energy spectrum of the Mini-X2 divided into $50$ channels.

\item Detector of \SI{1}{cm^3} placed at \SI{50}{cm} distance from the source as shown in Figure \ref{fig:Dosimentro a distanza 50 cm e fascio GPS}.

\item 120° isotropic source placed at the origin. The {\it General Particle Source} was used, a point-like source that allows particle emission for a given solid angle.
\end{itemize}

\begin{figure}[htbp]
\centering
\includegraphics[width=1\textwidth]{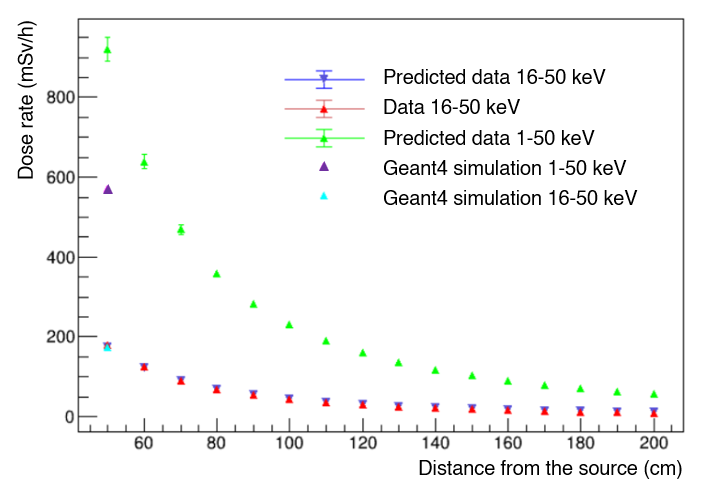}
\caption{Comparison of the predicted dose rate, the experimental measurements and the simulated values predicted by Geant4.}
\label{fig:Confronto teorica esperimento e simulazione}
\end{figure}

From the simulation both the dose rate calculated taking into account only the energy channels to which the dosimeter is sensitive to and the total dose rate due to the entire energy spectrum were considered. In Figure \ref{fig:Confronto teorica esperimento e simulazione} these two distributions are shown, along with the predicted distribution calculated using (\ref{eq:3.14}) and the experimental data from which we are able to make some considerations.

\begin{figure}[htbp]
\centering
\includegraphics[width=1\textwidth]{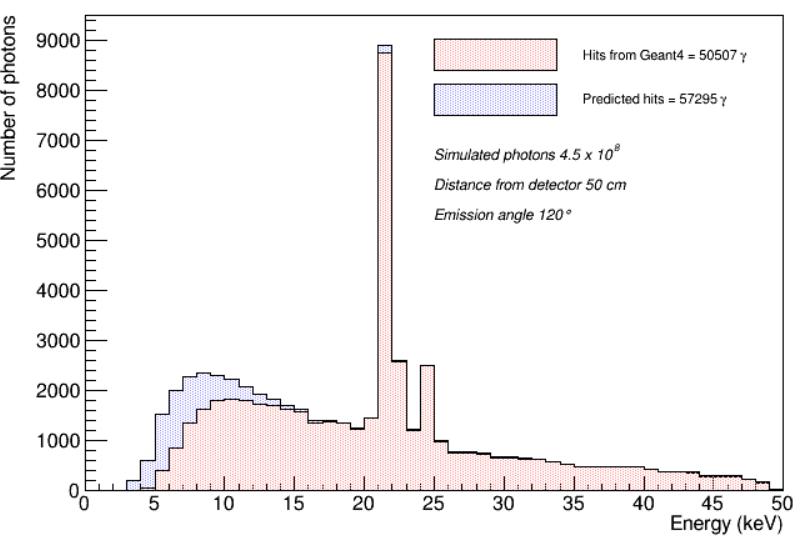}
\caption{Geant4-Theory comparison of the number of photons hitting a detector at \SI{50}{cm} distance. The integral of the histogram corresponds to the total number of photons in the respective sample.}
\label{fig:Confronto Teo Geant numero fotoni}
\end{figure}

First, it is observed that if the first 15 energy channels are excluded, the predicted dose rate (blue dots, Figure \ref{fig:Confronto teorica esperimento e simulazione}) is in very good agreement with the experimental one (red dots, Figure \ref{fig:Confronto teorica esperimento e simulazione}). However, when considering all 50 channels of the spectrum, the predicted data (green dots, Figure \ref{fig:Confronto teorica esperimento e simulazione}) deviates considerably from the experimental one (red dots, Figure \ref{fig:Confronto teorica esperimento e simulazione}). This is due to the less energetic photons that, interacting more easily in air, arrive at a lower percentage at the dosimeter. These interactions are not considered in the theoretical calculation, which consequently returns a higher value. The light blue dot and the purple dot are dose rate values obtained by Geant4 simulations only at \SI{50}{cm} from the source. We can see how the first one, obtained excluding the first 15 channels, is perfectly superimposable to the respective red and blue dot. It means that Geant4 reproduce in a good approximation the dose rate measured by the dosimeter. Finally, we had Geant4 return the number of photons that, for each energy channel, hit the dosimeter. We then compared this value (Figure \ref{fig:Confronto Teo Geant numero fotoni}) with the expected number calculated as 

\begin{equation} 
\label{eq:3.29}
\phi_{E,r} = \frac{\phi_{E}}{\pi r^2}
\end{equation}
where $\phi_{E,r}$ is the number of photons with energy $E$ striking the dosimeter at distance $r$ while $\phi_{E}$ is the number of photons with energy $E$ emitted from the source.

As hypothesized, from about \SI{15}{keV}, the two histograms (Figure \ref{fig:Confronto Teo Geant numero fotoni}) are perfectly superimposable while below this value the number of hits deviates. It turns out that about 

\begin{equation} 
\label{eq:3.30}
1 - \frac{50507}{57295} = 0.1185 = 11.85 \%
\end{equation}
of photons do not reach the detector. However, because they have energy below the sensitive range of the dosimeter used, this cannot be verified experimentally. Finally, it is interesting to note that although the number of particles not arriving is about 10\%, it contributes to a dose rate reduction of about 30\%. This is due to the exponential trend of the mass-energy attenuation coefficient that appears in (\ref{eq:3.14}). 

\paragraph{Shielding construction}
\label{sec:intro}

In order to be able to work with the Mini-X2, the construction of adequate shielding around the X-ray tube was essential. The large angle of emission and the possibility of scattering do not permit use without shielding. The purpose of such shielding is therefore to allow the operator to work close to the source without being exposed to the outgoing radiation. 

The development of shielding around the Mini-X2 must take into account several factors. First is the optimization between effectiveness, production cost and transportability. Therefore, materials already available were used, trying to optimize them as mentioned above. Specifically, the materials were:

\begin{itemize}
\item Two aluminum sheets: one of thickness \SI{3}{mm} and one of thickness \SI{1}{cm}
\item A sheet of copper \SI{1.5}{mm}-thick 
\item A sheet of lead \SI{2}{mm}-thick 
\end{itemize}

For each of these materials we tabulate their atomic number Z and the value of the mass-energy attenuation coefficient for the most penetrating photons of \SI{50}{keV}.

\begin{table}[htbp] 
    \centering  
    \caption{Atomic number and $\mu_{en}/\rho$ of Al, Cu, Pb}  
    \smallskip
    \begin{tabular}{c|c|c}
    \hline
       \textbf{Element}  &  \textbf{Z}  &  $\bm{\mu_{en}}$/$\bm{\rho}$ $\bm{(cm^2/g)}$ \\
       \hline
       Al  &  $14$  &  1.840E-01 \\
       Cu  &  $29$  &  2.192E+00 \\
       Pb  &  $82$  &  6.740E+00 \\
       \hline
    \end{tabular}   
    \label{tab:4}
\end{table}

\begin{figure}[htbp]
\centering 
\subfigure[] 
  {\includegraphics[width=0.47\textwidth]{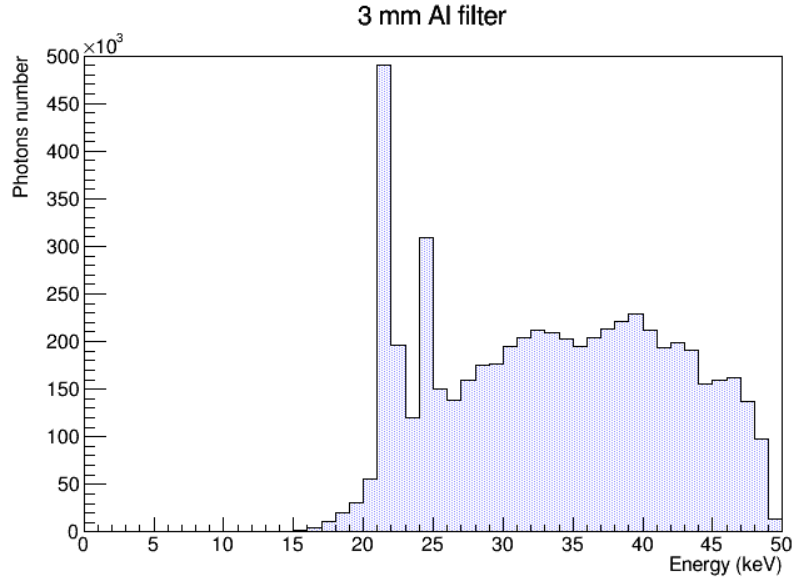}\label{fig:imm}}
\hspace{5mm} 
\subfigure[] 
  {\includegraphics[width=0.47\textwidth]{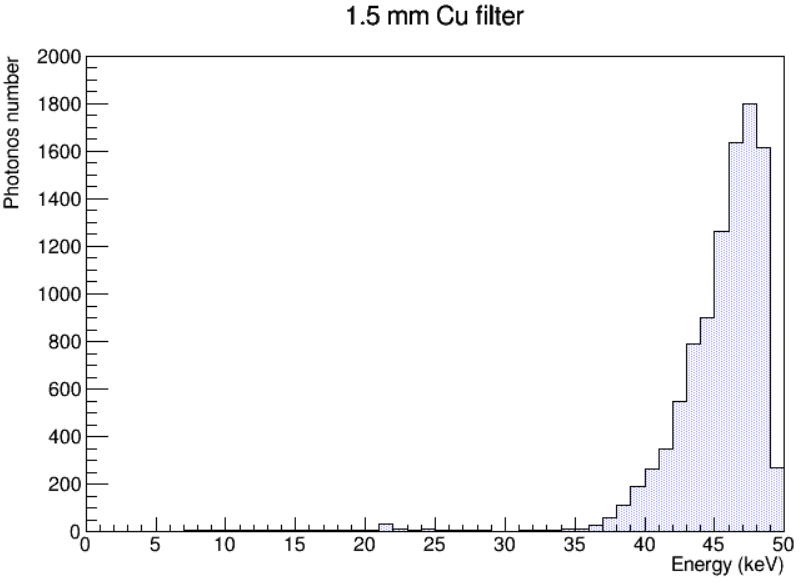}}
  \label{}
\subfigure[] 
  {\includegraphics[width=0.47\textwidth]{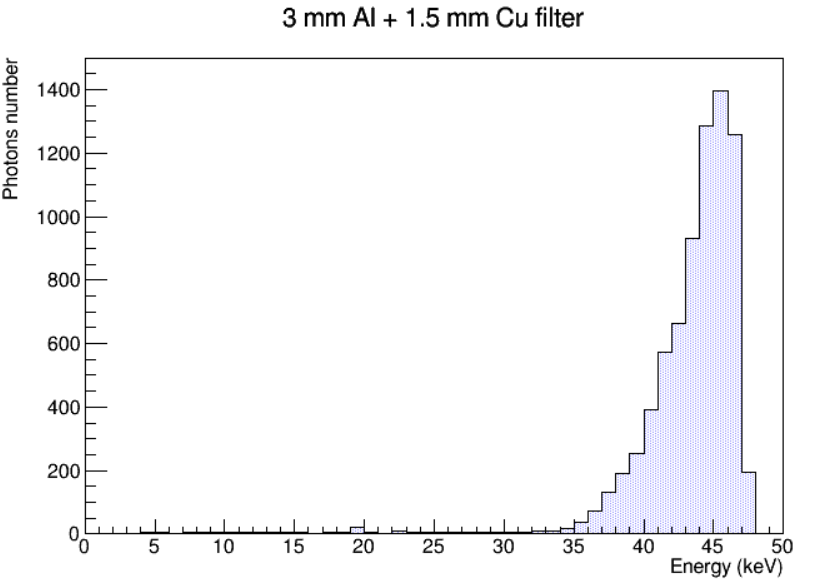}}
\hspace{5mm}
\subfigure[] 
  {\includegraphics[width=0.47\textwidth]{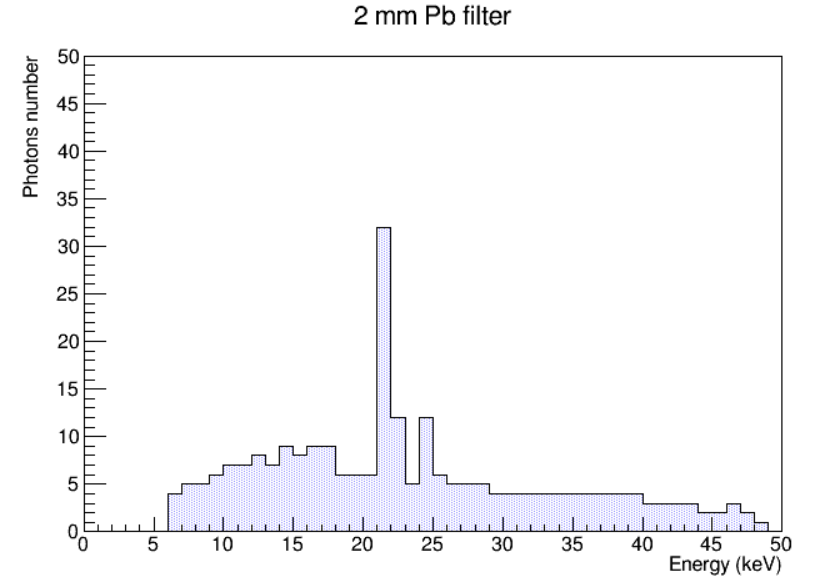}}
\caption{Energy spectra resulting from the application of different filters to the Mini-X2. The integral of the area corresponds to the total number of photons that manage to pass through the filters.}
\label{fig:Spettri con i filtri}
\end{figure}

For each of the materials, it is possible to know their shielding effect by simulating with Geant4 a setup consisting of a source emitting $4.5 \cdot 10^7$ photons against a screen of \SI{1}{m^2} in front of a detector of the same size.

\begin{figure}[htbp]
\centering
\includegraphics[width=1\textwidth]{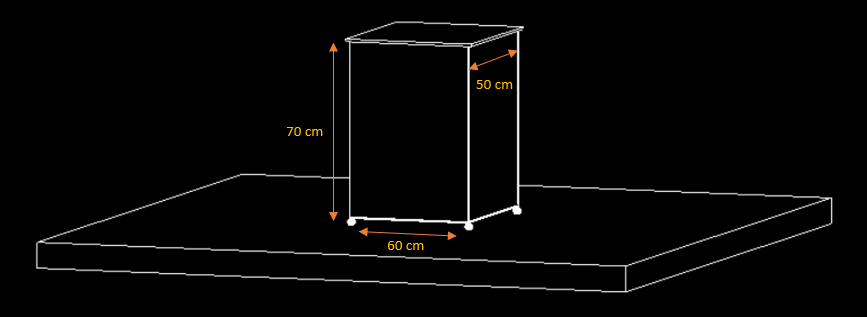}
\caption{Simulated box.}
\label{fig:Box simulata}
\end{figure}

Observing in Figure \ref{fig:Spettri con i filtri} how the spectrum is modified by the individual Al, Cu and Pb shielding, it becomes apparent how it is necessary to layer several materials to obtain acceptable protection. The only material that individually succeeds in providing good protection is lead, which reduces the spectrum by a factor of $10^7$. As far as Al and Cu are concerned, the shielding effect of using them together turns out to be satisfactory. In this case only the most energetic photons survive, which, however, due to the low mass-energy attenuation coefficient, contribute to a minimum dose value.  Based on this information, a shielding box was simulated around Mini-X2 composed of at least \SI{3}{mm} of Al plus \SI{1.5}{mm} of Cu. The dimensions of the box must then take into account the fact that we need to accomodate inside a GEM detector of \SI{10}{cm} x \SI{10}{cm}. Therefore, the dimensions shown in Figure \ref{fig:Box simulata} were chosen.

As for the screens these were made up as follows: 

\begin{itemize}
\item On the side faces by an inner layer of \SI{1.5}{mm} of Cu plus an outer layer of \SI{3}{mm} of Al
\item On the upper face by an inner layer of \SI{1.5}{mm} of Cu plus an outer layer of \SI{1}{cm} of Al 
\item On the lower face by an inner layer of \SI{2}{mm} of Pb plus an outer layer of \SI{3}{mm} of Al 
\end{itemize}

The presence of lead in the lower face is not accidental: in fact, orienting the Mini-X2 downward requires more reinforcement of the lower layer. The upper face has also been strengthened, this is because, given the height of the box, a hypothetical person close to the box is more exposed to this side and to those photons that after scattering with the inner walls manage to exit the upper screen. In addition, a configuration in which the heaviest material is on the inside was purposely chosen to reduce scattering. In fact, for the same photon energy, the photoelectric effect is favored in a heavier material. During this process the photon is absorbed resulting in the emission of an electron. This electron, if directed outwards, would cross the outermost screen. Since the electron will have low energy, it will tend to lose all its energy by ionization. If photons were to cross the lighter material, the Compton effect would be favored. This would cause some of the photons to reach the outermost (denser) layer, where the photoelectric effect would be dominant. Therefore, in this configuration electrons emitted outwards would not be shielded.

%If this were the innermost layer, then some of the photons would reach the outermost surface after scattering. In this one the photoelectric effect would be favored, however, this time the electrons emitted outward would not be shielded by additional layers.

\paragraph{Simulations with Person}
\label{sec:intro}

The dose absorbed by a person is obtained through the formula

\begin{equation} 
\label{eq:3.17}
D = \frac{dE_{dep}}{dm}
\end{equation}
therefore, unlike the simulations carried out so far, it is now necessary to determine from Geant4 the value of energy deposited in a given volume of mass $m$. Recalling that  

\begin{equation} 
\label{eq:3.18}
m = \rho V
\end{equation}
it is possible to simulate a person of about \SI{75}{kg} by constructing a volume of size

\begin{figure}[htbp]
\centering
\includegraphics[width=0.8\textwidth]{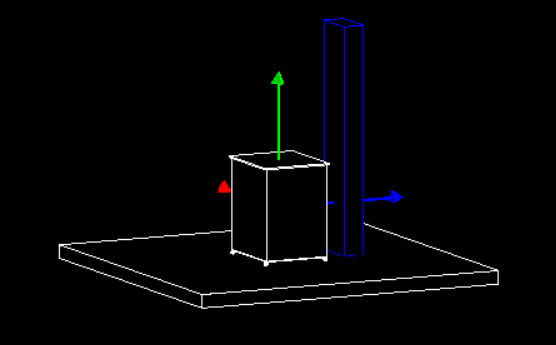}
\caption{Simulation setup with a person positioned at 15 cm from the side of the box}
\label{fig:Setup persona e box}
\end{figure}

\SI{170}{cm} in height, \SI{33}{cm} in front width and \SI{14}{cm} of length as shown in Figure \ref{fig:Setup persona e box} and consisting of water. 

\begin{figure}[htbp]
\centering
\includegraphics[width=0.8\textwidth]{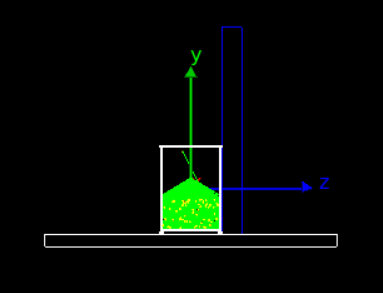}
\caption{Side view of the simulation setup with the person next to the box side and the source active.}
\label{fig:Setup persona e box di profilo}
\end{figure}

With regard to the simulations carried out, as mentioned in the previous chapters, the highest risk conditions were simulated. Therefore, the parameters entered within Geant4 are:

\begin{itemize}
\item Number of photons related to the maximum applicable current (in accordance with an acceptable computational time), namely $4.5 \cdot 10^8 \hspace{0.1cm} \gamma/s$.
\item A person in very close proximity to the side of the box with a width of \SI{60}{cm}, in order to be as close as possible to the source.
\item Source positioned as high as possible. Considering the dimensions of the Mini-X2, this means placing the source \SI{8.8}{cm} above the center of the box.
\end{itemize}

\begin{figure}[htbp]
\centering
\includegraphics[width=0.9\textwidth]{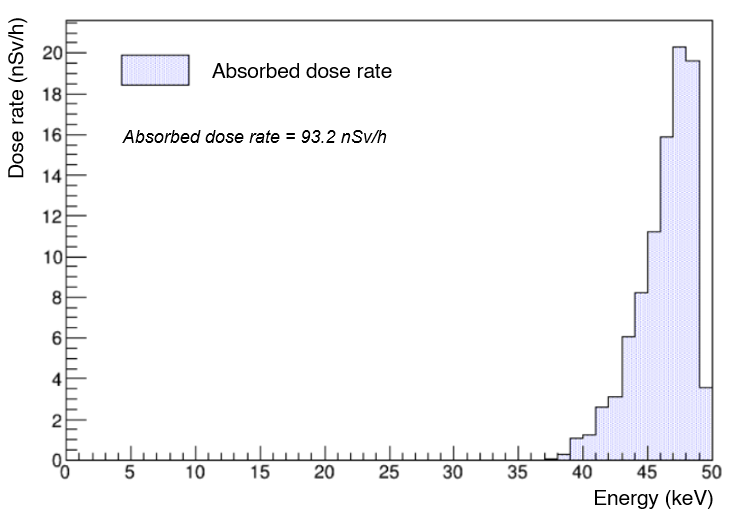}
\caption{Absorbed dose rate by a person at a distance of \SI{0}{m} from the box.}
\label{fig:Dose assorbita dalla persona a 0 m}
\end{figure}

\begin{figure}[htbp]
\centering
\includegraphics[width=1\textwidth]{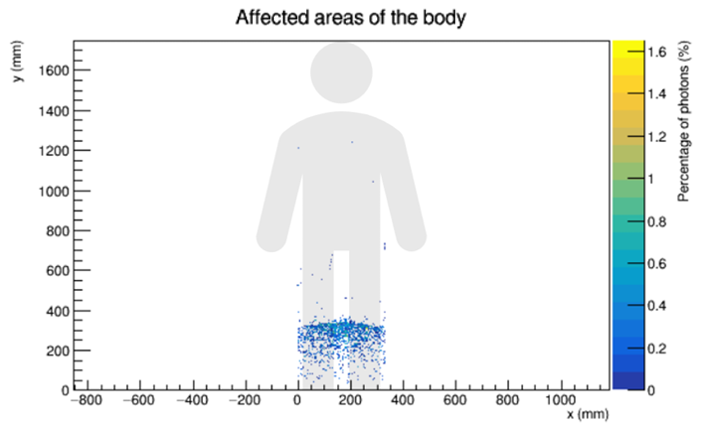}
\caption{Affected points of the body.}
\label{fig:Punti del corpo colpiti}
\end{figure}

From Geant4, the absorbed dose rate is obtained, along with the body parts that are affected by the few particles that manage to penetrate the shielding. In Figure \ref{fig:Dose assorbita dalla persona a 0 m} the distribution of the dose absorbed by the person is reported. In Figure \ref{fig:Punti del corpo colpiti}, the points hit by the outgoing particles are schematically shown on an XY plane. The clear separation that forms around at $30-35$ cm of height can be explained by referring to the Figure \ref{fig:Setup persona e box di profilo}: the photons that manage to survive are mostly those that undergo no scattering and reach the walls of the box directly. On the other hand, photons that bounce inside the box progressively lose energy and fail to pass through the shielding.

\begin{figure}[htbp]
\centering
\includegraphics[width=0.5\textwidth]{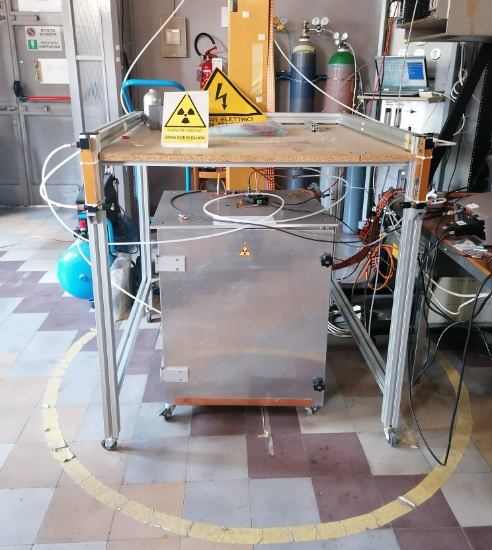}
\caption{Photo of the box inside the monitored area.}
\label{fig:Area sorvegliata}
\end{figure}

Since this value is just slightly above natural radiation level, the absorbed dose result is satisfactory. In reality, the experiments were conducted under safer conditions. In fact, since the experimental setup in the laboratory forces us to remain at a certain distance from the box, we were never at a 0 distance from it. (Figure \ref{fig:Area sorvegliata}).

However, simulations were carried out in other conditions. The resulting dose rate values are reported in Table \ref{tab:5}, only the absorbed dose rate values\footnote{The dose rate in the table takes into account the transition from the simulated $10^8$ photons to the real $10^{11}$ photons.}

\begin{table}[htbp] 
    \centering  
    \caption{Dose rate simulations for different operating conditions} 
    \smallskip
    \begin{tabular}{c|c|c|c}
    \hline
       \textbf{Photons number}  &  \textbf{Current}  &  \textbf{Distance from the box}  & \textbf{Dose rate} \\
       \hline
       $4.5 \cdot 10^8$ $\gamma$/s  &  $200$ $\mu$A  &  $50$ cm &  $5.1$ nSv/h  \\
       $1.8 \cdot 10^8$ $\gamma$/s  &  $80$ $\mu$A  &  $0$ cm &  $36.9$ nSv/h    \\
       \hline
    \end{tabular}   
    \label{tab:5}
\end{table}
After obtaining these results, the shielding described above was constructed. A site inspection by a qualified expert then verified the operating conditions of the device and authorized its use.

\section{Experimental analysis of Mini-X2 X-ray Tube}
\label{sec:intro}

In this section, we will show the confirmation of the simulation studies using experimental measurements. For this type of verification, the idea is to place a dosimeter at the center of the lower face of the box, take the dose rate value, and compare it with a simulation of the same setup. Subsequently, the linearity between the current applied to the Mini-X2 and the number of photons emitted was analyzed. In this case as well, theoretical data were compared with experimental data.

\paragraph{Experimental setup}
\label{sec:intro}

\begin{figure}[htbp]
\centering 
\subfigure[The open box is shown in profile view. Inside, the Mini-X2 is oriented downward at the maximum possible height. At the bottom of the box, the dosimeter is positioned, oriented 90° relative to the direction of the source.] 
  {\includegraphics[width=0.4\textwidth]{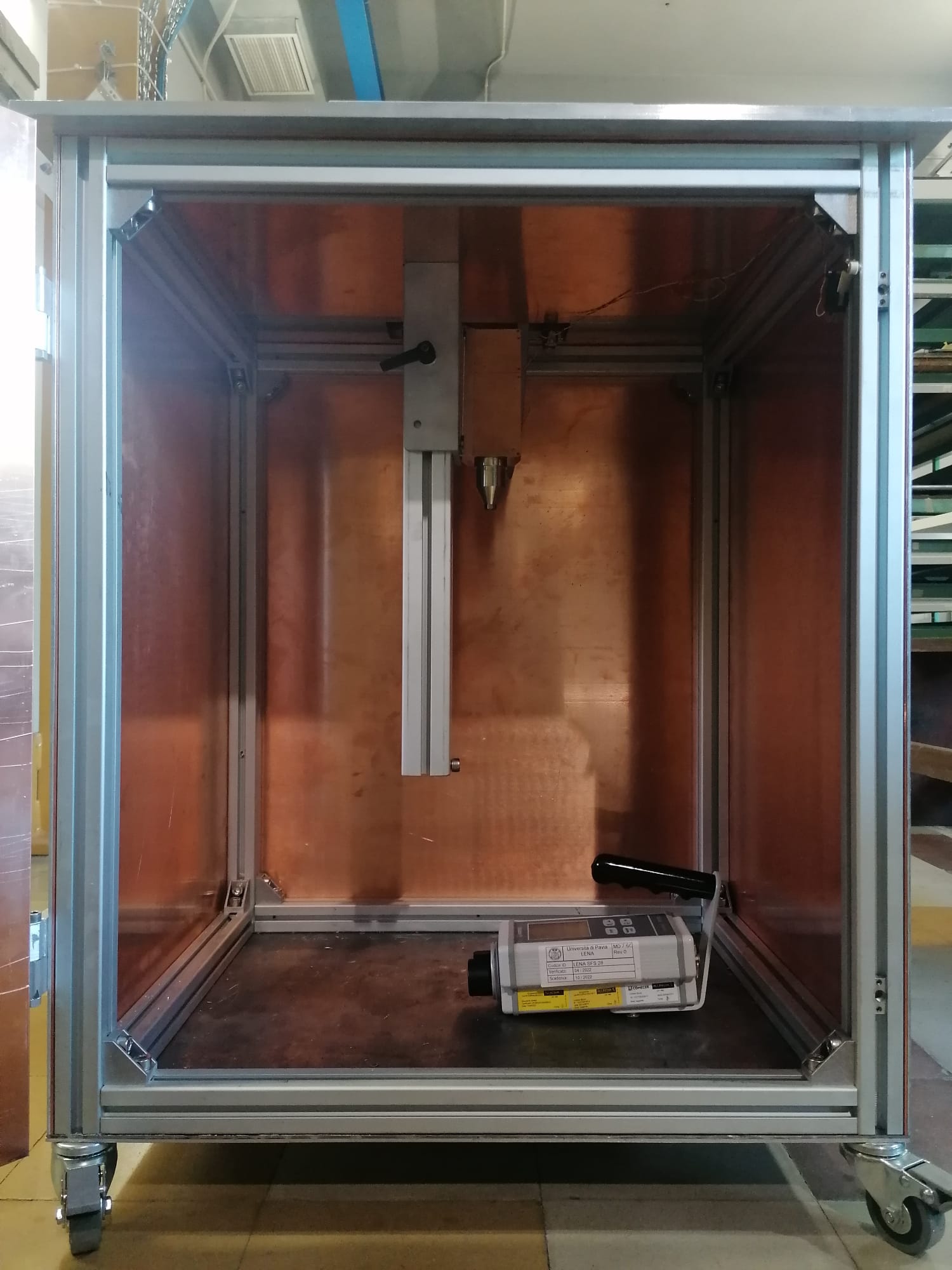}\label{fig:Dosimetro orrizzontale in box}}
\hspace{5mm} 
\subfigure[Top part of the box. There are a series of connectors for the GEM cables, an LED, and the Mini-X2 controller.] 
  {\includegraphics[width=0.4\textwidth]{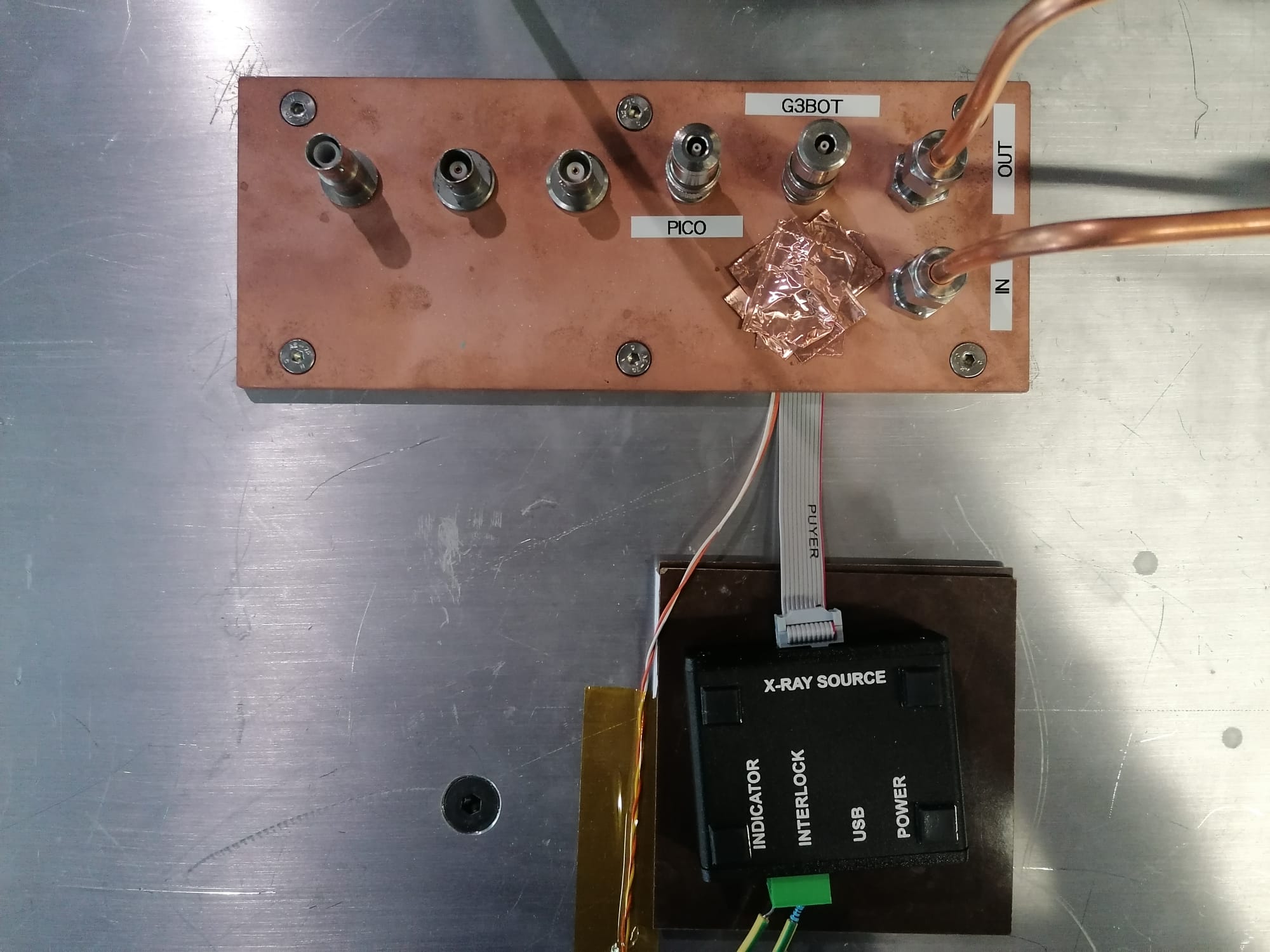}\label{fig:Cavi box}} 
\caption{} 
\label{fig:Cavi e dosimetro in box}
\end{figure}

The construction of the box included not only additional precautions related to shielding protection but also user convenience. For example, as shown in the Figure \ref{fig:Dosimetro orrizzontale in box} a support was mounted inside to allow movement of the Mini-X2 along the vertical axis, from a maximum distance from the bottom of \SI{46.2}{cm} to a minimum of \SI{14.3}{cm}. The Figure also shows the relative positions between the source and the detector for the measurements mentioned in the previous section. In particular, the distance between the Mini-X2 and the dosimeter cap is \SI{41.8}{cm} $\pm$ \SI{0.1}{cm}.

Regarding the additional precautions, the setup was equipped with LED, which lights up when the Mini-X2 is turned on, and an interlock system. The latter is a safety system connected to the box closure that allows the source to be turned on from the PC only if the box is sealed.

Finally, for safety reasons, the box was placed inside an already existing monitored area in the laboratory, marked off with tape. This required a minimum distance of at least 30 cm from the walls of the box.

\paragraph{Dosimeter operation}
\label{sec:intro}

\begin{figure}[htbp]
\centering
\includegraphics[width=0.30\textwidth]{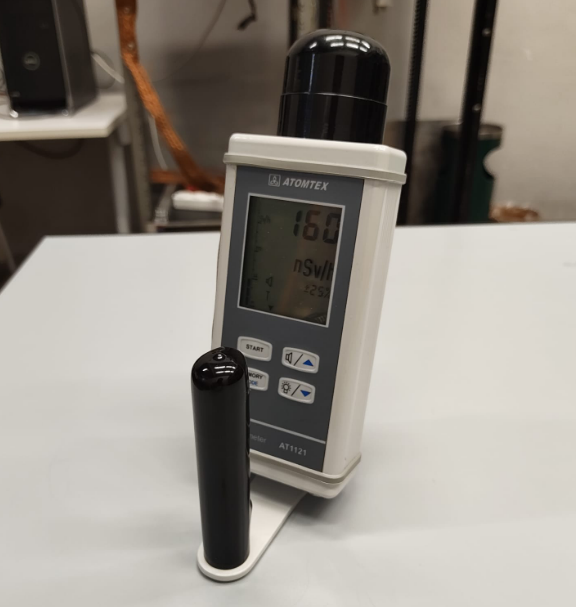}
\caption{Dosimeter used.}
\label{fig:Dosimetro utilizzato}
\end{figure}

For the experimental measurements, the AT1123 dosimeter shown in Figure \ref{fig:Dosimetro utilizzato} from the ATOMTEK manufacturing company was used~\cite{ref10}~\cite{ref11}.

Its operation is based on a scintillator and a photomultiplier. Furthermore, it is capable of detecting only photons with energies greater than \SI{15}{keV}. This meant it isn't possibile directly compare the dose value read on the dosimeter with the value obtained from the simulations. Therefore, in the comparison between simulations and experimental measurements, this cutoff must be taken into account, and only data from \SI{15}{keV} and above should be compared.

An additional feature that received attention is the different response of the dosimeter when oriented at various angles relative to the source.

\begin{table}[htbp]
  \centering
  \caption{Response in the case of the source above the dosimeter at different angles and for photons at different energies~\cite{ref11}.}
  \begin{tabular}{>{\centering\bfseries}c|*{10}{c}}\hline
    
    \multirow{2}{*}{\bfseries Energy (keV)} & \multicolumn{10}{c}{\bfseries Degree, incidence of radiation}  
    \\\cline{2-11}
    & \textbf{0} & \textbf{+15} & \textbf{+30} & \textbf{+45}
    & \textbf{+60} & \textbf{+75} & \textbf{+90} & \textbf{+105}
    & \textbf{+120} & \textbf{+135} \\ \hline
    
    22.0 & 1.00 & 0.96 & 0.90 & 0.85 & 0.80 & 0.70 & 0.60 & 0.50 & 0.40 & 0.05 \\ \hline
    
    59.5 & 1.00 & 1.00 & 0.95 & 0.93 & 0.90 & 0.85 & 0.80 & 0.75 & 0.50 & 0.30 \\ \hline
    
    662.0 & 1.00 & 1.00 & 1.00 & 1.00 & 1.00 & 0.97 & 0.95 & 0.93 & 0.90 & 0.80 \\ \hline

    1250.0 & 1.00 & 1.00  & 1.00 & 1.00 & 1.00 & 1.00 & 1.00 & 0.96 & 0.95 & 0.85 \\ \hline
    
  \end{tabular}
  \label{tab:15}
\end{table}

As noted in the previous section, the dosimeter inside the box is oriented at 90° relative to the direction of the Mini-X2. Therefore, its response undergoes a variation, which is described in the table \ref{tab:15}. Considering that the spectrum of the Mini-X2 has a peak at \SI{22}{keV}, for each measurement taken with the experimental setup described in the previous section, a reduction factor approximately equal to $0.6$ of the actual value must be considered.

Finally, the dosimeter is capable of storing the maximum dose rate value from the moment the START button is pressed to begin the measurement. This way, it can be placed inside the box while the source is active, and the dose rate can be measured.

\paragraph{Experimental verification of Mini-X2}
\label{sec:intro}

\begin{figure}[h]
\centering
\includegraphics[width=1\textwidth]{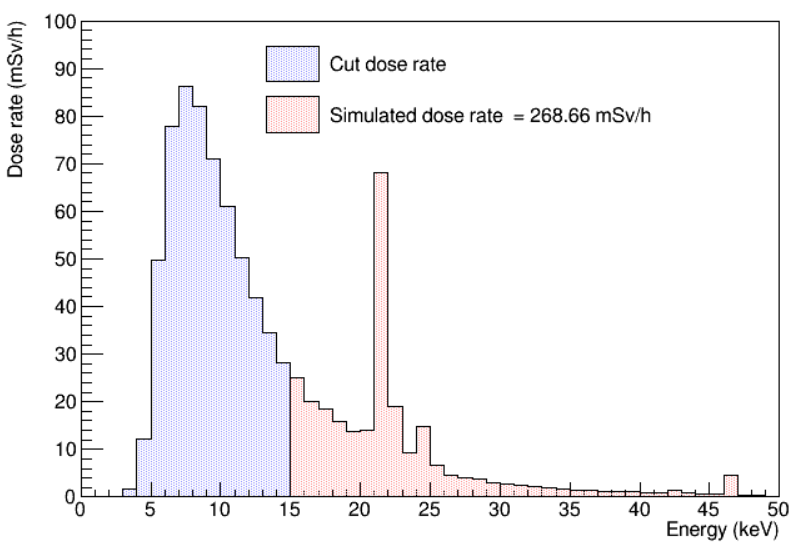}
\caption{Simulation of the dose rate measured by the dosimeter (red) and of the dose rate due to the contribution of all channels (blue). It includes the rescaling from $10^7$ to $10^{11}$ $\gamma$/s, but not the rescaling due to the orientation of the dosimeter..}
\label{fig:Sim dose dal dosimetro}
\end{figure}

To experimentally compare the Mini-X2 with theoretical data, a simulation was performed under the conditions described by the experimental setup. The result is shown in Figure \ref{fig:Sim dose dal dosimetro}. We simulated the following features:

\begin{itemize} 
\item A total number $4.5 \cdot 10^7 \hspace{0.1cm} \gamma/s$ associated with a current of $\SI{200}{\mu A}$ 
\item Dosimeter placed along the axis of the Mini-X2 at a distance of \SI{41.8}{cm} from it 
\item Isotropic emission angle of 120°. 
\end{itemize}

However, given the operation of the dosimeter and the number of simulated photons, the experimental value does not match the simulated one. It is necessary to first rescale from $10^7$ to $10^{11}$ $\gamma$/s, obtaining the dose rate 

\begin{equation} 
\label{eq:3.19}
\dot D_{sim} = 268.66 \pm 0.57 \hspace{0.2cm} mSv/h
\end{equation}
where the error is calculated as $\sigma/\sqrt{N}$ since three independent simulations were performed.

Subsequently, remembering the orientation of the dosimeter, to obtain the hypothetical experimental data, it is necessary to multiply by $0.6$ obtaining

\begin{equation} 
\label{eq:3.20}
\dot D_{sim} = 161.29 \pm 0.34 \hspace{0.2cm} mSv/h
\end{equation}

At this point three experimental measurements were carried out, obtaining the data in table \ref{tab:6}, thanks to which we obtain
\begin{equation} 
\label{eq:3.21}
\dot D_{exp} = 159.33 \pm 4.78 \hspace{0.2cm} mSv/h
\end{equation}
The data is in perfect agreement with the simulation and this confirms all the work done previously.

\begin{table}[h]  
    \centering
    \caption{Dose measurements inside the box. The 1\% error is taken from the dosimeter manual, which assigns this uncertainty whenever the dose rate is on the order of mSv/h.} 
    \smallskip
    \begin{tabular}{c|c|c}
    \hline
       \textbf{Current}  &  \textbf{Distance from box}  & \textbf{Dose rate}\\
       \hline
       $200$ $\mu$A  &  $41.8$ cm &  $(161.00 \pm 1.61)$ mSv/h \\
       $200$ $\mu$A  &  $41.8$ cm &  $(158.00 \pm 1.58)$ mSv/h \\
       $200$ $\mu$A  &  $41.8$ cm &  $(159.00 \pm 1.59)$ mSv/h \\
       \hline
    \end{tabular}   
    \label{tab:6}
\end{table}

\paragraph{Linearity between applied current and emitted photons}
\label{sec:intro}

The linearity between current and photons has already been introduced. In this section, it will be demonstrated, through experimental analysis, that

\begin{equation} 
\label{eq:3.22}
\gamma_{emitted} = K \cdot i
\end{equation}
where $K$ is a proportionality constant and $i$ is the current applied to the X-ray tube. The constant $K$ appearing in (\ref{eq:3.22}) depends on the experimental setup used, as it is related to the number of photons that reach the dosimeter.

As a theoretical hypothesis, it is convenient to start from the calculated data (\ref{eq:3.20}) simulated in the previous section:

\begin{equation} 
\label{eq:2.10}
\dot D_{200_{\mu A}} = 161.29 \pm 0.34 \hspace{0.2cm} mSv/h
\end{equation}
Thanks to this, it is possible, assuming linearity and a fixed distance, to theoretically hypothesize the dose rate value at different currents. Then, for the same current values, experimental measurements were taken, yielding the data shown in table \ref{tab:7}.

\begin{table}[htbp]   
    \centering  
    \caption{Predicted and experimental dose rate at differt currents.\label{tab:7}}
    \smallskip
    \begin{tabular}{c|c|c}
    \hline
       \textbf{Current} & \textbf{Predicted dose rate} & \textbf{Experimental dose rate}  \\
       \textbf{($\bm{\mu}$A)} & \textbf{(mSv/h)} & \textbf{(mSv/h)} \\
       \hline
       $200$   &  $161.29$   &  $158$    \\
       $180$   &  $145.16$   &  $145$    \\
       $160$   &  $129.03$   &  $127$    \\
       $140$   &  $112.90$   &  $111$    \\
       $120$   &  $96.77$    &  $95$     \\
       $100$   &  $80.65$    &  $80$     \\
       $80$    &  $64.52$    &  $64$     \\
       $60$    &  $48.39$    &  $48$     \\
       $40$    &  $32.26$    &  $32$     \\
       $20$    &  $16.13$    &  $15.7$   \\
       $5$     &  $4.03$     &  $3.4$    \\
       \hline
    \end{tabular}  
\end{table}

\begin{figure}[h]
\centering
\includegraphics[width=1\textwidth]{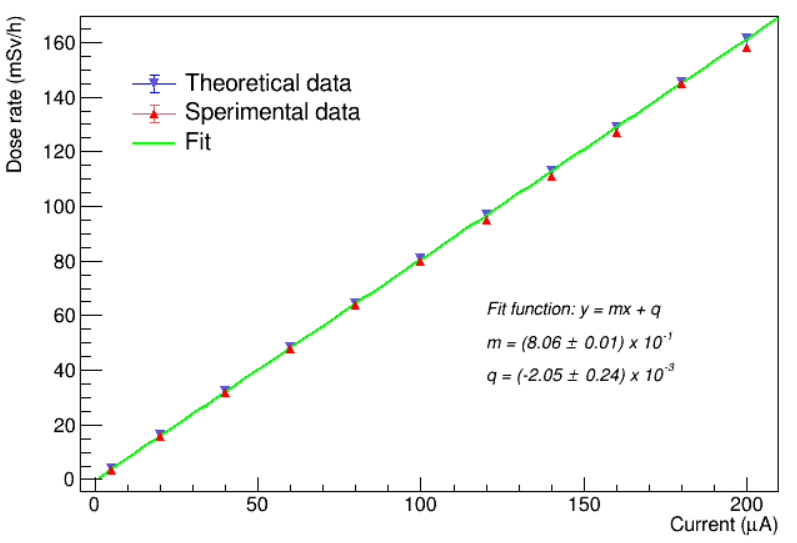}
\caption{Comparison between theoretical data and experimental data at different operating current of the Mini-X2.}
\label{fig:Linearità corrente fotoni Spe vs Teo}
\end{figure}

By plotting the tabulated values along with the errors and fitting the curve, an excellent approximation between the theoretical and experimental data is obtained. (Figure \ref{fig:Linearità corrente fotoni Spe vs Teo}).

From the fit it is clear that there is a direct proportionality between the dose rate and the current, given by the formula

\begin{equation} 
\label{eq:3.24}
\dot D = 0.807 \cdot i
\end{equation}
Since, from (\ref{eq:2.13}), the dose rate is directly proportional to the number of photons emitted, the sought linearity is therefore demonstrated. In particular, considering the conditions under which the measurements were taken, we have

\begin{equation}\begin{split}
\label{eq:3.25}
\dot D_{air} &= 5.76 \cdot 10^{-4} \frac{\phi}{\pi \cdot 41.8^2}\sum_{i=16}^{50} k_i \hspace{0.06cm} E_i \hspace{0.06cm} \left(\frac{\mu_{en}}{\rho}\right)_i\\
&= 5.57 \cdot 10^{-10} \hspace{0.1cm} \phi
\end{split}\end{equation}
in which

\begin{equation} 
\label{eq:3.26}
\phi = 1.45 \cdot 10^9 \hspace{0.1cm} i
\end{equation}
where the factor $1.45 \cdot 10^9$ has units of [1/C] and $i$ is in $\mu A$. It is important to note that this formula applies only inside the box, so it includes the factor $0.6$ due to the dosimeter's orientation and an additional factor related to those photons that would normally not hit the detector but reach it due to scattering on the walls, thus increasing the dose value and consequently the value of $\phi$. In fact, if one wanted to calculate the number of photons emitted at $\SI{200}{\mu A}$ using (\ref{eq:3.26}) and accounting for the dosimeter orientation, the formula would be

\begin{equation} 
\label{eq:3.27}
\phi_{200_{\mu A}} = \frac{1.45 \cdot 10^9}{0.6} \hspace{0.1cm} \cdot 200 = (4.83 \pm 0.06) \cdot 10^{11} \hspace{0.2cm} \gamma/s
\end{equation}
which, as can be seen, is a slightly higher value compared to the actual $4.5 \cdot 10^{11} \hspace{0.2cm} \gamma/s$. This confirms that the proportionality constant $K$ introduced earlier depends on the setup. Finally, thanks to this analysis, it is possible to determine the percentage of additional photons that reach the dosimeter due to scattering when it is inside the box.

\begin{equation} 
\label{eq:3.28}
\frac{4.83 \cdot 10^{11}}{4.5 \cdot 10^{11}} = 1.073 
\end{equation}
which corresponds to about $7.3$\% more photons.

\section{Conclusion}
\label{sec:intro}

In conclusion, the main results obtained during this work are summarized. The Mini-X2 X-Ray Tube was characterized by estimating both its emissive power and its dependence on the tube current. Based on these measurements, the absorbed dose rate as a function of distance was theoretically predicted. To enable experimental comparison and ensure safe operation, these theoretical values were used to design a shielding system capable of attenuating the X-ray beam, using a series of simulations performed with \textit{Geant4}. Finally, the dose rate values inside the shielding box—obtained theoretically—were compared with those from \textit{Geant4} simulations and with experimental measurements carried out using a dosimeter and the values obtained in these differents methods are in good agreement. In this configuration the linearity between current and number of emitted photons was proven.


\begin{thebibliography}{}

\bibitem{ref1} D. Ravanelli, \textit{Valutazione della dose assorbita dal paziente in radiologia digitale: sviluppo di un modello fisico basato sull’imaging digitale}, 2008, pp. 15--61.

\bibitem{ref2} F. H. Attix, \textit{Introduction to Radiological Physics and Radiation Dosimetry}, 2004, pp. 26, 125--142, 206, 528.

\bibitem{ref3} XrayConsult, “Nota tecnica di radioprotezione,” \url{https://www.xrayconsult.it/nota-tr.html}, accessed May 2023.

\bibitem{ref4} C. H. Hahn \textit{et al.}, “First Observation of Signals Due to KAERI’s 10 MeV Electron Beam by Using GEM Detectors,” \textit{Journal of the Korean Physical Society}, vol. 50, no. 4, Apr. 2007, p. 967.

\bibitem{ref5} Radiopaedia, “X-ray quantity and quality,” \url{https://radiopaedia.org/articles/x-ray-quantity-and-quality}, accessed May 2023.

\bibitem{ref6} M. Bhat, J. Pattinson, G. Bibbo, M. Caon, “Diagnostic X-ray spectra: A comparison of spectra generated by different computational methods with a measured spectrum,” Oct. 29, 1997, p. 116.

\bibitem{ref7} Amptek, “Mini-X2 X-Ray Tube,” \url{https://www.amptek.com/products/mini-x2-x-ray-tube}, accessed May 2023.

\bibitem{ref8} NIST, “X-Ray Transition Energies Database,” \url{https://physics.nist.gov/PhysRefData/XrayTrans/Html/search.html}, accessed May 2023.

\bibitem{ref9} NIST, “X-Ray Attenuation – Mass Attenuation Coefficient of Air,” \url{https://physics.nist.gov/PhysRefData/XrayMassCoef/ComTab/air.html}, accessed May 2023.

\bibitem{ref10} Atomtex, “AT1121, AT1123 X-Ray and Gamma Radiation Dosimeters,” \url{https://atomtex.com/en/at1121-at1123-x-ray-and-gamma-radiation-dosimeters}, accessed May 2023.

\bibitem{ref11} Atomtex, \textit{Manuel d'utilisation ATOMTEK AT1121}, \url{http://www.laprev.fr/images/Manuel-dutilisation-AT1123.pdf}, accessed May 2023.

\bibitem{ref12} CERN, “Geant4 Software,” \url{https://geant4.web.cern.ch/}, accessed May 2023.



\end{thebibliography}
\end{document}